\def\VeriPyArxiv{1}
\documentclass{article}
\PassOptionsToPackage{numbers,sort&compress}{natbib}
\usepackage[final]{neurips_2026_vericode}
\usepackage[utf8]{inputenc}
\usepackage[T1]{fontenc}
\usepackage{amsmath,amssymb}
\usepackage{graphicx}
\usepackage{booktabs,tabularx}
\usepackage{xcolor}
\usepackage{listings}
\usepackage{tikz}
\usepackage{microtype}
\usepackage{url}
\usepackage{hyperref}
\hypersetup{hidelinks,pdftitle={VeriPy Source-Preserving Verification and Compatibility Checking for Python Components},pdfauthor={Naing Oo Lwin}}
\definecolor{vpySkyBlue}{HTML}{72BEF2}
\definecolor{vpyGreen}{HTML}{0F734F}
\definecolor{vpyLavender}{HTML}{8B6DC5}
\definecolor{vpyDeepPurple}{HTML}{48267F}

\colorlet{vpyInk}{vpyDeepPurple!35!black}
\colorlet{vpyMuted}{vpyDeepPurple!70!white}
\colorlet{vpyLine}{vpyDeepPurple!24!white}
\colorlet{vpyBlueFill}{vpySkyBlue!22!white}
\colorlet{vpyGreenFill}{vpyGreen!10!white}
\colorlet{vpyLavenderFill}{vpyLavender!16!white}
\colorlet{vpyPurpleFill}{vpyDeepPurple!8!white}

\definecolor{vpyDafnyBlue}{HTML}{184275}
\definecolor{vpyDafnyYellow}{HTML}{FFE75C}

\lstdefinestyle{vpycode}{
  basicstyle=\ttfamily\footnotesize\color{black},
  keywordstyle=\bfseries\color{black},
  commentstyle=\color{black},
  stringstyle=\color{vpyGreen},
  numbers=left,numberstyle=\tiny\color{black!55},
  numbersep=5pt,xleftmargin=13pt,
  showstringspaces=false,keepspaces=true,columns=fullflexible,
  breaklines=true,breakatwhitespace=false,
  postbreak={},
  aboveskip=4pt,belowskip=4pt,
  tabsize=4,upquote=true
}
\lstdefinelanguage{Dafny}{
  morekeywords={lemma,if,then,else,function,predicate,var,int,string,bool},
  morekeywords=[2]{requires,ensures,assert,decreases},
  keywordstyle=[2]\color{vpyDeepPurple},
  morecomment=[l]{//},
  morestring=[b]"
}
\newcommand{\examplemark}[1]{%
  \tikz[baseline=(label.base)]\node[draw=black,text=black,
  circle,inner sep=1.2pt,font=\sffamily\scriptsize] (label) {#1};%
}

\lstdefinelanguage{VeriPyPython}[]{Python}{
  deletecomment=[l]{\#},
  morekeywords=[2]{ensures,ghost_ensures,invariant,decreases,proof},
  keywordstyle=[2]\color{vpyDeepPurple}
}

\newcommand{\toolname}{VeriPy}
\newcommand{\code}[1]{\begingroup\def\_{\textunderscore\allowbreak}\texttt{#1}\endgroup}
\ifdefined\VeriPyArxiv
\title{\raisebox{-0.3em}{\includegraphics[height=1.4em]{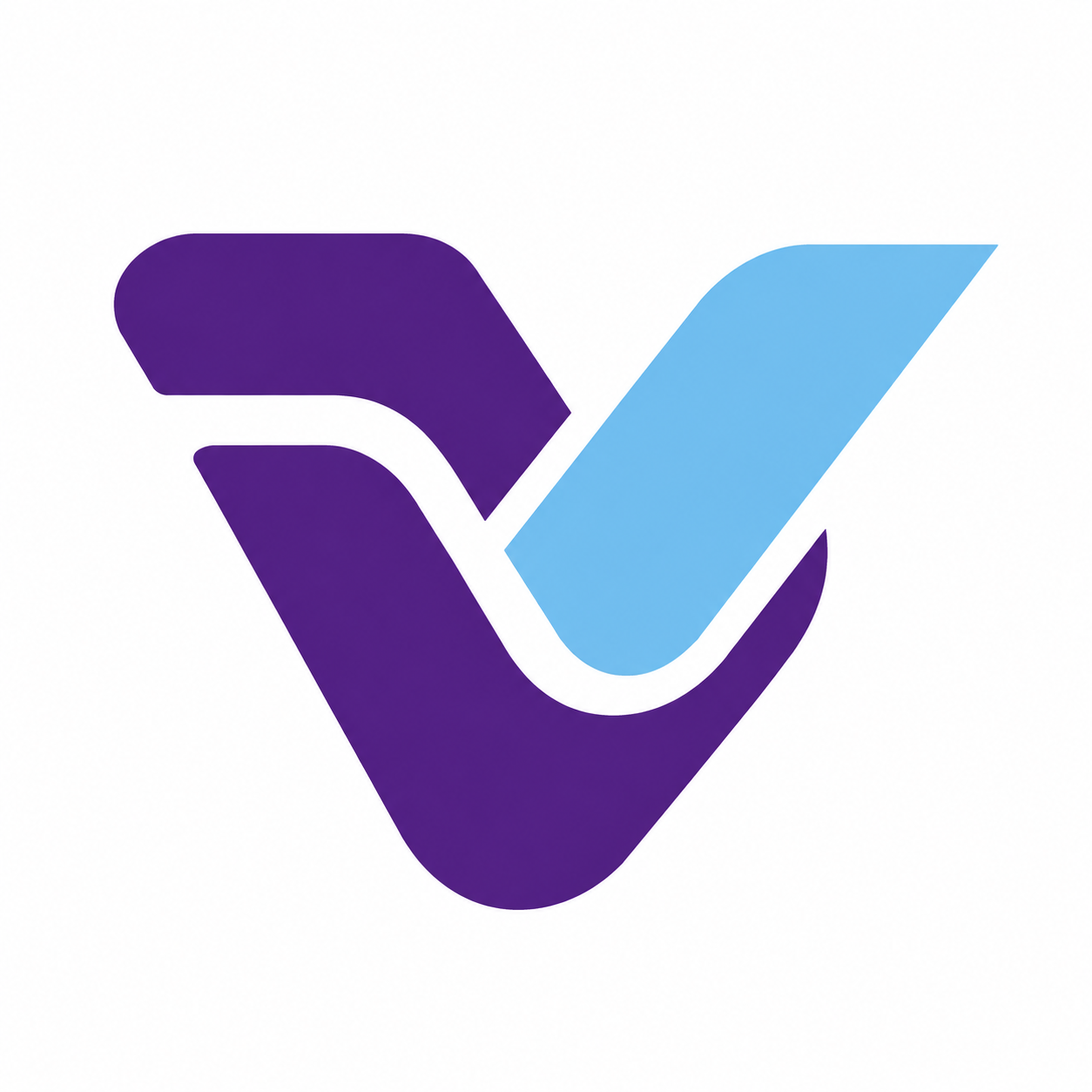}}\,VeriPy Source-Preserving Verification and\\Compatibility Checking for Python Components}
\else
\title{VeriPy Source-Preserving Verification and\\Compatibility Checking for Python Components}
\fi
\workshoptitle{AI for Verifiable Coding (VeriCodeGen)}
\author{Naing Oo Lwin\\Astrio Labs\\\texttt{nolan.lwin@astriolabs.com}}
\begin{document}
\maketitle

\begin{abstract}
Keeping a Python program and its formal guarantees aligned is a continuing
maintenance problem. Specifications must describe the implementation that
actually runs, and updates must preserve the behavior on which existing callers
depend. VeriPy brings these obligations into a common workflow for annotated
Python components. Developers and agents express contracts, invariants, and
proof hooks as Python comments, then refine checked auxiliary lemmas using
source-located diagnostics. Direct encoders produce Dafny or Lean artifacts
while retaining admitted executable bodies and explicit models of their
dependencies. A relational product checks whether an updated component still
admits old inputs and preserves returned values and modeled exceptions.
The resulting workflow connects source-level proof development with functional
verification and backward compatibility, recording the assumptions under which
each guarantee applies. The code is available at
\url{https://github.com/astrio-labs/veripy}.
\end{abstract}

\section{Introduction}
\label{sec:introduction}

Software maintenance changes the context in which a correctness guarantee must
hold. A component may be verified against a contract and later replaced by an
implementation that satisfies the same contract but behaves differently for its
callers. Even before an update, the proof must account for the executable body,
its dependencies, and the runtime assumptions under which the model applies.
For Python, these concerns connect two verification problems. The first relates
an existing component to its formal specification. The second relates successive
versions through the behavior they expose.

Source-oriented verification provides a foundation for this connection.
Frama-C supports C contracts through ACSL \citep{kirchner2015framac,framacwp},
and Nagini verifies annotated Python through Viper
\citep{eilers2018nagini,mueller2016viper}. Verus separates executable,
specification, and proof code in Rust \citep{lattuada2023verus}.
Strata provides extensible analysis infrastructure in Lean, including Python
analysis through its Laurel intermediate language \citep{strata2026architecture}.
These systems establish the value of keeping specifications close to programs.
Our focus is the connection between proof development, preserved Python
components, and compatibility with their earlier versions.

\toolname{} makes the Python component the developer-facing unit of verification.
Contracts, loop invariants, termination measures, and proof hooks appear as
\texttt{\#@} comments alongside executable statements. Developers and agents use
source-located diagnostics to refine annotations and auxiliary lemmas in checked
sidecars. A typed admission gate and explicit dependency models determine the
behavior represented by separate Dafny and Lean encoders. Backend checking
establishes the generated obligations while admitted executable bodies remain
available to native callers. Optional guards enforce supported input conditions.
The proof report records the environment and assumptions needed to interpret
the result as a guarantee about that component.

Compatibility adds an obligation that functional verification alone does not
supply. A range postcondition, for example, allows multiple valid outputs and
therefore cannot establish agreement between versions. \toolname{} constructs a
relational product that checks admission of every old input and equality of
returned values and modeled exception outcomes. Exact functional contracts and
checked relation lemmas support this comparison. Reports distinguish a proved
relation, a concrete native witness, an inconclusive attempt, and an unsupported
boundary, so failure to construct or discharge a proof is not reported as a
behavioral difference.

This paper contributes an implemented workflow connecting source-level proof
assistance, explicit Python models, and scoped compatibility checking.
Django's base36 rewrite demonstrates the complete path from checked component
contracts to a product proof across different algorithms. Repository studies
examine proof support and native integration, while controlled maintenance
trials separate old-proof access from available library information. Frozen
history searches also expose unsuccessful transfer to unseen changes.
Together, these results characterize what the workflow supports and where
preparation and semantic coverage still limit automation.

\section{VeriPy by example}
\label{sec:example}

\begin{figure}[!b]
\centering
\begin{minipage}[t]{.54\linewidth}
\vspace{0pt}
{\small\bfseries\color{black}Python implementation}
\begin{lstlisting}[style=vpycode,language=VeriPyPython,escapeinside={(*@}{@*)}]
#@ ensures raised() <==> old(i) < 0
#@ ghost_ensures not raised() ==> ghost("ExactEncoding", old(i), result) (*@\examplemark{A}@*)
def int_to_base36(i: int) -> str:
    char_set = '0123456789abcdefghijklmnopqrstuvwxyz'
    if i < 0:
        raise ValueError('Negative base36 conversion input.')
    if i < 36:
        return char_set[i]
    b36 = ''
    while i != 0:
        #@ invariant i >= 0
        #@ invariant len(b36) > 0 or i >= 36
        #@ invariant ghost("EncodingProgress", old(i), i, b36) (*@\examplemark{B}@*)
        #@ decreases i (*@\examplemark{C}@*)
        #@ proof EncodingStep(old(i), i, b36, char_set) (*@\examplemark{D}@*)
        i, n = divmod(i, 36)
        b36 = char_set[n] + b36
    return b36
\end{lstlisting}
\end{minipage}\hfill
\begin{minipage}[t]{.42\linewidth}
\vspace{0pt}
{\small\bfseries\color{black}Checked proof step}
\begin{lstlisting}[style=vpycode,language=Dafny,numbers=none,xleftmargin=0pt]
lemma EncodingStep(original: int, remaining: int, suffix: string, chars: string)
  requires remaining > 0
  requires EncodingProgress(original,remaining,suffix)
  requires chars == "0123456789abcdefghijklmnopqrstuvwxyz"
  ensures EncodingProgress(original,remaining/36,[chars[remaining%36]]+suffix)
{
  assert Digits(remaining) == (if remaining < 36 then [chars[remaining%36]] else Digits(remaining/36)+[chars[remaining%36]]);
}
\end{lstlisting}
\end{minipage}
\caption{\textbf{A real Django compatibility example.}
The revised base36 converter retains every executable statement and annotation
from the checked Python view. Only its docstring is omitted. Circled letters
mark annotations explained in the text. The right panel
shows the checked Dafny lemma called before the quotient/remainder update.
The full old implementation
appears in Appendix~\ref{app:old-code}.}
\label{fig:django-example}
\end{figure}

Django's \texttt{int\_to\_base36} is our principal changed-code compatibility
example. Password-reset-token code uses its output to encode a timestamp. The old algorithm emits digits from left to right. The later one uses
\texttt{divmod} and prepends digits. They must agree on exact strings for
nonnegative integers and \texttt{ValueError} for negative inputs.

Figure~\ref{fig:django-example} shows the later implementation with the
annotations used in our recorded comparison. The executable statements are
preserved. The historical extraction specializes the comparison to Python~3
and removes an unreachable Python~2 branch. This adaptation is recorded rather
than treated as byte-for-byte preservation of the enclosing module.

The developer first fixes an exact-\texttt{int} boundary and checks admission.
The retained development logs initially reject parameter rebinding and mixed
tuple assignment. These required compiler work, not stronger annotations.
After admission, a return-site diagnostic reports ``a postcondition could not
be proved on this return path'' (source line 16, related contract line 2).
This identifies an unproved obligation, not a counterexample or inferred repair.
The developer supplies the encoding predicate, progress invariant, and checked
lemmas shown below. The recorded final functional check succeeds.

Let $D(n)$ be the canonical base36 encoding of $n\geq 0$, defined recursively by
the checked sidecar. The postconditions at \examplemark{A} specify exceptional
behavior and name
$\mathit{ExactEncoding}(n,s)\equiv n\geq0\land s=D(n)$.
The invariant at \examplemark{B} uses \texttt{EncodingProgress} to relate
the original input $n$,
the remaining quotient $r$, and suffix $s$. Here $D(n)=s$ when $r=0$, and
$D(n)=D(r)\mathbin{+\!\!+}s$ otherwise, where $+\!\!+$ concatenates strings.
The \texttt{\#@ decreases}
annotation at \examplemark{C} supplies a termination measure. The
\texttt{\#@ proof} line at \examplemark{D} instantiates
\texttt{EncodingStep} before the next executable update. Dafny checks the
lemma's body and its call preconditions.
Neither annotation executes in CPython.

The old implementation has separate invariants and checked support. The
product checks old-input admission and equal outcomes. The exact encoding
predicate uniquely determines normal results, and negative inputs have
matching exception outcomes. Rechecking the retained Dafny product for this paper completed with
267 verified obligations and no errors. This count describes generated
obligations, not independent programs or test cases.

For deployment, a separate Django 6.0 study inserts the guarded converter into
its real token caller and retains six timestamp comparisons with identical
tokens, alongside 1,957 native/guarded/compiled cases. These bounded checks
connect the component to a caller. They do not prove authentication security
or compatibility of that Django release. The historical algorithm proof and
this integration study use distinct pinned sources. Annotation and sidecar
work remains manual or agent-assisted. Historical authoring time is unknown.

\section{Verification workflow and semantic boundary}
\label{sec:system}

\subsection{Admission and explicit environments}

\begin{figure}[!t]
  \centering
  \includegraphics[width=\linewidth]{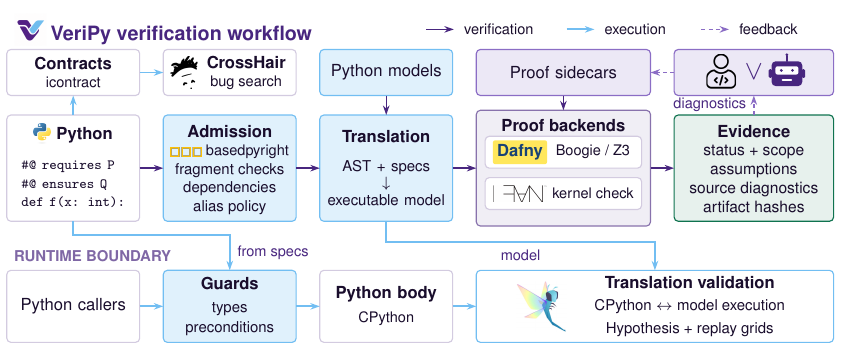}
  \caption{\textbf{VeriPy verification workflow.}
  VeriPy parses Python and \texttt{\#@} specifications, checks types with
  basedpyright, and admits a supported fragment. Dafny or Lean checks the
  translated program and any proof sidecars supplied by a human or LLM.
  CrossHair searches executable contracts for violations. Guards enforce
  admitted input conditions. Generated guards and the generic Hypothesis
  harness target Dafny. Case-study replay coverage varies by backend.
  Execution checks compare the original body and the translated model
  independently of proof success. Verification reports record assumptions
  and provenance. Solid purple arrows show verification dependencies.
  Solid blue arrows show generation, execution, and testing dependencies.
  Dashed lavender arrows show the proof-assistance loop.}
  \label{fig:veripy-verification}
\end{figure}

Figure~\ref{fig:veripy-verification} separates proof dependencies, execution
checks, and proof assistance. The input is a Python component together with its
annotations and a declared environment. basedpyright checks its static types
\citep{basedpyright}. Structural admission then determines whether the selected
backend supports the syntax and operations. Type checking does not establish
functional correctness, and a supported type does not imply that every Python
operation on that type is admitted.

The admitted domains use exact builtin values or explicitly modeled snapshots.
This matters because a Python subclass can override an operation that the model
treats as builtin. Source preservation is checked at the executable-AST level
after erasing type annotations. Stronger interface assumptions, collection
specialization, and extraction adaptations are recorded separately.
For example, Black's general collection interfaces are specialized to builtin
lists in the verified views. For framework objects, adapters construct snapshots
of the fields used by the component. Stable reads during snapshot construction
remain an assumption.

Module-level constants are resolved by a bounded AST evaluator over admitted
initialization forms. The evaluator can use preceding same-module helper
definitions without executing arbitrary initializers. Dependencies require an
explicit model. An import statement by itself grants no proof authority.
The retained environment records constants, supported dependency operations,
exception hierarchies, and any registered regular expressions. Selected
integration adapters check source, proof, and dependency hashes before loading
a saved guard, so changed inputs require a proof refresh.

String operations use an explicit Python/Unicode configuration. The extension
studies admit Unicode scalar strings and reject lone surrogates at their
boundaries. Decimal conversion accounts for the configured digit limit.
Regular-expression support covers a registry of the studied Packaging and
Werkzeug patterns, rather than arbitrary patterns. These choices make the
supported behavior inspectable and allow unsupported operations to be rejected
before they acquire an unjustified model.

\subsection{Specifications, translation, and proof checking}

The annotation language includes preconditions, postconditions, bounded
quantification, loop invariants, termination measures, and named proof hooks.
Outcome specifications distinguish normal return from modeled exceptions.
Expressions can refer to the result and to entry-state values.
A ghost predicate such as \texttt{ExactEncoding} in
Section~\ref{sec:example} is defined and checked in the sidecar. It is not an
assumed property of the Python function.

The implementation has separate direct AST encoders for Dafny and Lean,
with shared admission and dependency machinery. It does not introduce a
backend-neutral intermediate language. The Dafny path emits programs and
contracts for SMT-backed verification \citep{leino2010dafny}. The Lean path
emits executable definitions and propositions whose proofs are checked by
Lean's kernel \citep{demoura2021lean}. Backend support is not identical.
In particular, the historical relational checker currently targets Dafny,
and the generic guard and differential-test generators use Dafny-oriented
lowerings. Dedicated case-study replays cover the Lean artifacts.

Proof sidecars are regenerated into complete verification artifacts.
The Dafny sidecar gate restricts declarations and rejects explicit assumptions
and axiomatic additions. Lean sidecars are checked for permitted declarations
and forbidden ambient-context commands, and exported theorems receive axiom
audits. The allowed logical axioms are distinguished from new user assumptions.
These gates are part of the trusted implementation, not themselves a verified
security boundary.

An audit exposed why checking a backend's exit code or axiom list is
insufficient. A Lean sidecar could introduce an ambient hypothesis of type
\texttt{False}. The kernel then checked a conditional theorem, while the
frontend reported the intended unconditional contract as proved. The fix rejects
such ambient declarations and retains an adversarial reproducer. A proof report
must correspond to the requested proposition, including its hypotheses, as well
as to an accepted backend artifact.

\subsection{Execution, guards, and the trust boundary}

Generated guards enforce supported exact-type, dependency, alias, and
precondition checks around the Python body. Some integrations copy inputs into
snapshots. In-place byte-buffer studies instead retain the original buffers and
check permitted identities before mutation. For PyPNG, the result may alias the
scanline as the real reader requires, while the previous row must remain
distinct. A universal copy-in policy would change the behavior being integrated.

Guard generation consumes source and specifications and is independent of proof
success. A guard can reject an inadmissible call or check a supported
postcondition at runtime, but its existence is not evidence that the component
was verified. Conversely, comment annotations and proof hooks add no execution
to the unwrapped function. Guard checks and adapters have measurable costs.

Empirical checks compare Python observations with executable translations.
The generic harness uses Hypothesis \citep{maciver2019hypothesis}. Individual
studies also use deterministic grids, upstream tests, independent oracles, and
kernel-reduced Lean executions. CrossHair can search executable contracts for
violations \citep{crosshair}, but it is an auxiliary path, not the source of the
deductive results reported here. Matching executions do not prove translation
correctness.

A backend proof establishes a property of the translated model. Applying it to
Python relies on the frontend, semantic models, selected backend, and enforced
runtime assumptions. There is no mechanized end-to-end soundness theorem for
the complete encoder. Independent backend proofs and differential checks provide
distinct evidence, but shared frontend errors can affect both paths.

The intended assurance argument requires domain admission, corresponding
initialization and execution, preservation of specification hypotheses, and
binding of checked artifacts to the reported source and target. Compatibility
additionally requires old-domain inclusion and observational equality. The
artifact maps these obligations to concrete controls. Their complete conjunction
has not been mechanized.

\section{Scoped backward compatibility}
\label{sec:compatibility}

For version $v$, let $A_v(x)$ describe admission of an input $x$ under that
version's contracts and environment, and let $M_v(x)$ denote its translated
execution. Let $\mathcal{O}$ project an execution onto the declared observations.
For the terminating modeled computations in the evaluated boundary,
\toolname{} checks
\begin{equation}
\label{eq:compatibility}
  \forall x.\ A_{\mathrm{old}}(x)\ \Longrightarrow\
  \bigl(A_{\mathrm{new}}(x)\ \land\
  \mathcal{O}(M_{\mathrm{old}}(x))
  =\mathcal{O}(M_{\mathrm{new}}(x))\bigr).
\end{equation}
The admission implication prevents a proof obtained solely by strengthening
the new precondition. Normal observations are typed returned values. Exceptional
observations are modeled exception tags. Messages, tracebacks, object identity,
resource consumption, and concurrent interference are outside this relation.
Mutable-buffer poststates are an intended observation in the PyPNG study, but
the evaluated relational implementation does not yet admit that boundary.

The checker places independently encoded versions in separate namespaces and
constructs a sequential relational wrapper. It verifies both bodies, their
sidecars, and the wrapper together. This follows the established reduction of
relational properties to ordinary verification through product programs
\citep{barthe2011product}. It does not infer an alignment between arbitrary loops.
Exact functional contracts and, when necessary, a checked relation lemma provide
the facts needed at the two calls.

Matching parameter names, kinds, types, return types, and admitted immutable
literal defaults are required by the current implementation. Omitted arguments
are covered when those defaults match. Explicit environments are resolved
independently for the two versions. Exception names and modeled inheritance
must agree to compare their tags. Supported optional values preserve both
presence and payload. Changed defaults, variadic boundaries, unresolved
initialization, and mutable-buffer products remain unsupported.
Matching closed record schemas use a generated
field-preserving map $T$ between module-local datatypes. The product checks
$A_{\mathrm{new}}(T(x))$ and compares mapped returned values. This read-only
snapshot relation excludes identity and observable alias mutation.

\begin{figure}[t]
  \centering
  \includegraphics[width=\linewidth]{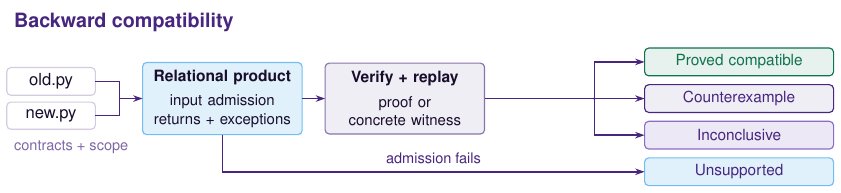}
  \caption{\textbf{Backward compatibility with VeriPy.} A relational product
  checks input admission, returns, and modeled exceptions over the old
  version's input domain. Compatibility is proved within the declared
  scope. Incompatibility requires a concrete witness. Admission failures
  return Unsupported before proof or replay. For admitted code, unsuccessful
  proof and replay without a witness remain Inconclusive. Arrows show
  possible control flow, not a successful verdict.}
  \label{fig:veripy-compatibility}
\end{figure}

Figure~\ref{fig:veripy-compatibility} shows the four reported outcomes.
\emph{Proved compatible} requires a successful product proof.
\emph{Counterexample} requires a concrete input with differing native
observations. \emph{Inconclusive} covers admitted comparisons whose proofs fail
or exhaust their budget without a witness. \emph{Unsupported} identifies a
boundary that the comparison pipeline cannot construct. Bounded replay is
additional evidence. Agreement cannot establish Equation~\ref{eq:compatibility},
and a replay contradiction invalidates a proof claim. Successful verification
of each version against a weak range contract is likewise insufficient.

\section{Evaluation}
\label{sec:evaluation}

We evaluate functional verification, historical compatibility, transfer to
previously unarchived changes, and runtime integration cost. Components were
selected for useful repository behaviors and developed interactively.
They are not a representative sample of Python projects.
Table~\ref{tab:historical} fixes one historical boundary per repository.
Broader functional closures, repeated helpers, and later refinements remain
separate evidence.

\subsection{Functional verification and backend coverage}

The dual-backend cohort contains 25 proof units across seven repositories and
32 repository/function-name identities, including evaluation helpers. Composed
units overlap and are not independent programs. The recorded Ubuntu CI checkpoint verified all
25 byte-identical annotated sources in Dafny 4.11.0 and Lean 4.33.1, yielding
50 successful invocations. Six group manifests match the frozen compiler.
Table~\ref{tab:parity} reports selected sizes, support and costs.
Appendix~\ref{app:effort} contains all units.

\begin{table}[t]
\caption{Component verification and support for nine selected units. Py denotes Python LOC.
Spec denotes contract LOC. Hint denotes inline proof-hint LOC. Support and time pairs are
Dafny / Lean. Support includes sidecar definitions and copied lemmas, excluding
shared imported libraries. All rows pass both backends. The complete 25-unit
inventory and counting rules are in Appendix~\ref{app:effort}.}
\label{tab:parity}
\centering\small
\begin{tabularx}{\linewidth}{@{}lrrrllX@{}}
\toprule
Component & Py & Spec & Hint & Support & Time (s) & Property \\
\midrule
Black range composition & 98 & 8 & 19 & 119 / 421 & 8.98 / 12.48 & Exact ordered output \\
SGLang live planner & 27 & 12 & 14 & 79 / 125 & 6.36 / 7.15 & Lengths and reserve \\
PyTorch shard validation & 94 & 18 & 27 & 357 / 1759 & 22.66 / 110.76 & Overlap and errors \\
Django base36 & 11 & 2 & 7 & 27 / 106 & 5.37 / 6.84 & Exact digits/errors \\
vLLM padding optimizer & 17 & 6 & 6 & 0 / 266 & 5.42 / 12.21 & Optimum and tie rule \\
CPython time parser & 37 & 34 & 0 & 0 / 976 & 12.73 / 163.59 & Exact fields/errors \\
Luhn checksum & 7 & 4 & 1 & 67 / 70 & 4.16 / 1.83 & Exact decimal sum \\
Luhn check digit & 13 & 12 & 2 & 120 / 260 & 7.49 / 2.14 & Decimal round trip \\
Luhn validation & 26 & 13 & 0 & 0 / 59 & 3.20 / 2.36 & Bounds/error classes \\
\bottomrule
\end{tabularx}
\end{table}

Additional Dafny extensions cover calendar helpers, name normalization, Luhn
validation, inverse PNG filters and ETag handling. The extension protocol required rerunning
all ten study pipelines at each recorded checkpoint. A completed full-suite checkpoint records 1,341 passing
regressions, which are engineering checks rather than additional case studies.
The later string-offset extension passes 339 focused tests and all fifteen
jobs across the ten studies, with no source drift.

Contract strength differs by component. CPython proves ordinal/date round trips
over unbounded positive years and ordinals through two new evaluation callers.
The original stdnum contracts establish errors, bounds and selected return
properties. Later overlapping refinements prove the exact decimal checksum and
check-digit round trip through both backends. These narrower results do not
replace the historical arbitrary-alphabet comparison.

CPython's time parser proves 34 postcondition clauses through both backends.
Its body is preserved with four type annotations, but Lean requires 976
nonblank, noncomment sidecar lines. This proves the helper's parsing behavior,
not ISO validation or historical parser equivalence.

Native and generated executions are compared at the declared callable boundaries.
Integration exercises public date calls, IMEI validation, complete PNG decodes,
and Response ETag operations. Appendix~\ref{app:reproduction} records the
separate workloads. Resolved initializer views and dependency stubs remain
explicit assumptions rather than evidence that all upstream annotations pass
strict typing.

\subsection{Historical compatibility}

\begin{table}[t]
\caption{Earlier compatibility checkpoint across ten pinned old/new boundaries. \emph{Changed} concerns the selected
executable body. Four proofs comprise two changed-code results and two
unchanged-code controls. Exact domains and revisions appear in
Appendix~\ref{app:domains}.}
\label{tab:historical}
\centering\small
\begin{tabularx}{\linewidth}{@{}lXcl@{}}
\toprule
Repository & Selected callable & Changed & Verdict \\
\midrule
Black & \code{is\_valid\_line\_range} & Yes & Proved \\
Django & \code{int\_to\_base36} & Yes & Proved \\
CPython & \code{\_days\_before\_year} & No & Proved \\
Werkzeug & \code{quote\_etag} & No & Proved \\
Packaging & \code{canonicalize\_name} & Yes & Counterexample \\
python-stdnum & \code{checksum} & No & Inconclusive \\
PyPNG & \code{undo\_filter\_paeth} & No & Unsupported \\
SGLang & \code{page\_aligned\_decode\_alloc\_lens} & No & Unsupported \\
PyTorch & \code{\_check\_shard\_metadata\_pair\_overlap} & No & Unsupported \\
vLLM & \code{\_approximate\_gcd} & No & Unsupported \\
\bottomrule
\end{tabularx}
\end{table}

Table~\ref{tab:historical} checks preservation between versions. Its two
changed-code proofs cover Black's redundant emptiness check on an exact
two-integer tuple and Django's different-loop algorithms, our principal
compatibility example (Section~\ref{sec:example}). CPython and Werkzeug are unchanged-body controls.
None establishes compatibility of the surrounding releases.

Packaging supplies three native witnesses. With validation enabled,
\code{name="foo\textbackslash n"} changes from a returned string to
\texttt{InvalidName}. Moving lowercase conversion before separator replacement
also changes Greek final sigma behavior. These domain-specific differences
need not be unintended defects.

stdnum remains inconclusive because range and error contracts do not determine
an exact arbitrary-alphabet checksum. PyPNG needs relational buffer mutation and
alias handling. The earlier SGLang and PyTorch boundaries lack record support,
and vLLM lacks an optional-parameter boundary.

Later frozen-record snapshots admit SGLang and PyTorch comparisons. SGLang adds
13 relation-proof LOC. PyTorch then passes under the frozen compiler without
edits or a relation sidecar. Both are unchanged-body controls. Four functional
checks pass across Dafny and Lean, while products use Dafny.
Appendix~\ref{app:records} reports effort and replay.

\subsection{Agent assistance and frozen coverage}

\begin{figure}[t]
\centering
\includegraphics[width=\linewidth]{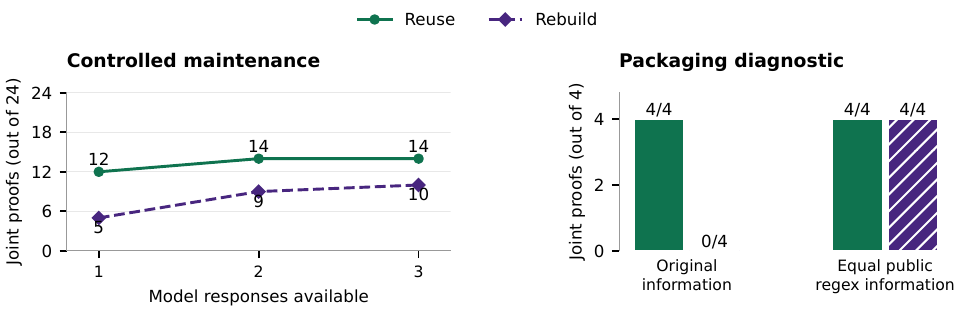}
\caption{Cumulative functional and compatibility proofs on twelve agent-authored
updates, repeated twice per arm. Repetitions are not independent samples.
The separate post-hoc Packaging diagnostic closes the gap with equal public
regex information, so these results do not isolate a proof-reuse benefit.
Appendix~\ref{app:maintenance-metrics} retains per-component counts.}
\label{fig:maintenance-completion}
\end{figure}

Figure~\ref{fig:maintenance-completion} examines whether access to checked old
proofs improves maintenance completion or response cost. The study seals proofs
for six human-reviewed components before generating twelve agent-authored
updates. Old functions contain 5 to 17 code lines. Updates are not upstream
commits, and no proposal is filtered by typing or proof success. Reuse receives
old proof annotations and helpers. Rebuild receives the fixed specification.
Both use \texttt{gpt-6-astra} with medium reasoning and identical sources,
contracts, public relations, checker and three-response budgets.
Four transformations have no private old proof support and are controls.

Reuse proves 14/24 trials (58.3\%) and rebuild 10/24 (41.7\%), representing
seven versus five distinct updates. First-response proofs are 12/24 versus
5/24. The completion gap occurs only in Packaging. On ten pairs where both
succeed, responses total twelve versus sixteen. Appendix~\ref{app:maintenance-metrics}
reports failure categories, token usage and measured time. These descriptive
costs depend on completion and stopping rules. Shared barriers
include three transformations with strict typing failures, one with unsupported
string augmented assignment, and one with ill-typed optional-string Dafny.

The old Packaging proof also exposes backend regex APIs. A separately frozen,
post-hoc diagnostic supplies the public regex definitions to both arms on the
same two updates. Both then pass all four trials in one response each. The
primary completion gap is sensitive to library information and does not isolate
a general lemma-reuse benefit.

A separate frozen history search selects three changes from twelve slots.
Under delegated model review, two Django proposals are rejected for old-invalid
behavior. An amended exploratory continuation proves one Werkzeug old component
after projecting typing overloads. All four maintenance trials remain unsupported.
A later calendar registration examines 134 path-history entries and selects no
eligible changed bodies, retaining both unfilled slots. Neither search supplies
a successful upstream transfer or a reuse-effect estimate.

Later string-affix development on the exposed Werkzeug update reaches
postcondition failures and a native whitespace witness. Native incompatibility
is witnessed, while the product-proof attempt remains inconclusive. This is not unseen transfer or a new agent trial.
Appendix~\ref{app:postfreeze-development} retains the development,
calendar readiness, review authority, and failures.

\subsection{Integration and runtime cost}

SGLang's guarded allocation planner was integrated into an installed EAGLE3
decode path. After correcting an overly restrictive pilot precondition, ten
matched fresh-process pairs run on one A10G at requested batch sizes 1, 8, 32,
and 128. Paired geometric-mean latency changes range from $-0.690\%$ to
$+0.839\%$. All family-adjusted 98.75\% bootstrap intervals include zero,
but negligible overhead is not established without a declared equivalence margin. Generated text
matches across 400 measured calls. Independent audits cover 1,920 planner
calls and 64,896 request entries.

Component guard costs can be substantial. Black's size-32 normalization takes
median native/guarded times of 3.22/69.51~$\mu$s. PyTorch constructor and metadata
checks also incur copying and deep-checking costs. Appendix~\ref{app:runtime}
reports workloads and uncertainty. These distinct experiments cannot be pooled
into a performance ranking.

\subsection{Negative controls and development effort}

Mutations change allocation, range, overlap, encoding or checksum behavior
while retaining the specifications. Native witnesses, backend failures,
timeouts and admission rejections remain distinct outcomes. Fresh Linux checks
pass 300 proof-boundary regressions and certify four concrete parser mutations
in the Lean kernel. An ordinary parser mutation run with mixed timeout and
error diagnostics is not counted as a clean rejection. Earlier audits cover
adversarial sidecars and 58 kernel-reduced executions of indexing, exceptions,
short-circuiting and nested control flow.

Development revises contracts, semantics and proofs interactively with AI
assistance. Its successful cases do not measure autonomous repair rates.
Sidecar sizes and selected tool times are recorded, but human hours and total
authoring model usage are unknown. We report no total development cost or
comparison of human proof effort between backends.

\section{Related work}
\label{sec:related}

\paragraph{Verification of mainstream languages.}
Frama-C integrates analyses of C programs around a common specification language
\citep{kirchner2015framac}. Its WP plugin proves ACSL contracts using weakest
preconditions and external provers \citep{framacwp}. Source annotations are
therefore an established verification interface.
Nagini combines static typing, contracts, and permission-based Python reasoning
through Viper \citep{eilers2018nagini,mueller2016viper}. It removes specification
calls before execution, so erasing proof annotations is also established.
\toolname{} targets narrower sequential components with comment annotations,
explicit dependency models, and native callers. We claim neither broader Python
coverage nor lower proof effort than Nagini.

Verus separates executable, specification, and proof code using Rust's ownership
discipline \citep{lattuada2023verus}. Aeneas translates supported Rust into
functional models \citep{ho2022aeneas}. Strata defines extensible intermediate
representations and analyses in Lean. Its Laurel intermediate language supports
analysis of Python and other imperative languages \citep{strata2026architecture}.
These systems establish source verification and reusable translation
infrastructure. \toolname{} focuses on preserving Python bodies across contracts,
native integration, and version relations, with explicit type and alias boundaries.
Its Dafny and Lean backends do not by themselves establish frontend correctness
\citep{leino2010dafny,demoura2021lean}.

\paragraph{Relational verification.}
Product programs reduce relational verification to a combined program
\citep{barthe2011product}. Frama-C's RPP plugin uses self-composition to verify
relations between C function calls \citep{blatter2018rpp}.
\citet{eilers2021wild} apply products in Nagini and show that separately sound
source encodings and product constructions need not compose soundly.
Our historical callable checks expose old-input admission and modeled exceptions.
The contribution is their implemented scope, verdict discipline, and repository
evidence, not a new relational calculus or a mechanized soundness result.

\paragraph{Testing and proof assistance.}
Hypothesis generates and reduces tests \citep{maciver2019hypothesis}, while
CrossHair searches for Python contract violations \citep{crosshair}.
\toolname{} combines these checks with proof sidecars and source-located feedback.
DafnyBench evaluates verification-hint completion \citep{loughridge2024dafnybench}.
DafnyPro combines executable-logic preservation, invariant pruning, and hint
augmentation \citep{banerjee2026dafnypro}. Our maintenance evaluation separates
preparation costs and preserves failed trials. Source-edit restrictions and
verifier feedback are established techniques, and comparative authoring
advantages remain unestablished.

\section{Limitations and implications}
\label{sec:limitations}

Component proofs depend on trusted semantic models, dependency resolution and
frontend translation. Runtime guards enforce only their implemented conditions.
Arbitrary objects, reflection, unrestricted regular expressions, resource
exhaustion and concurrent mutation remain outside scope. PyTorch overlap proofs
require positive shard widths, with native fallback for zero-size cases.
Black does not verify the native diff engine supplying match records.

Selected, manually annotated components and overlapping units limit generality.
GPU results cover one host/workload family with within-process drift. There is
no head-to-head verifier comparison.

Maintenance trials use small prepared functions and one model family. Public
API information affects completion. Archive novelty is not training-data
novelty. History searches produce no update proof, and calendar sampling finds
no eligible changes. Development on exposed cases is not unseen transfer.
Neither collaborator/model review nor finite examples certify specification
adequacy.

\section{Conclusion}

\toolname{} connects preserved Python components, checked contracts, native
callers, and scoped compatibility results. The studies establish useful
component capabilities while identifying preparation, semantic coverage, and
unseen upstream transfer as unresolved barriers to broader use.

\label{page:main-end}

\clearpage
\section*{Acknowledgments}
We thank Todd Schmid, Jules Jacobs, and Wojciech Różowski for their
advice and guidance, and Nada Amin and Robert Joseph George for early discussions.

\ifdefined\VeriPyArxiv\else
\section*{Funding disclosure}
This work received no funding.
\fi

\bibliographystyle{plainnat}
\bibliography{references}

@inproceedings{eilers2018nagini,
 author={Marco Eilers and Peter M{\"u}ller},
 title={{Nagini}: A Static Verifier for {Python}},
 booktitle={Computer Aided Verification (CAV)}, year={2018},
 pages={596--603}, doi={10.1007/978-3-319-96145-3_33},
 url={https://pm.inf.ethz.ch/publications/EilersMueller18.pdf}
}

@misc{lattuada2023verus,
 author={Andrea Lattuada and Travis Hance and Chanhee Cho and Matthias Brun and Isitha Subasinghe and Yi Zhou and Jon Howell and Bryan Parno and Chris Hawblitzel},
 title={{Verus}: Verifying {Rust} Programs using Linear Ghost Types (extended version)},
 year={2023}, howpublished={arXiv:2303.05491},
 url={https://arxiv.org/abs/2303.05491}
}

@inproceedings{mueller2016viper,
 author={Peter M{\"u}ller and Malte Schwerhoff and Alexander J. Summers},
 title={{Viper}: A Verification Infrastructure for Permission-Based Reasoning},
 booktitle={Verification, Model Checking, and Abstract Interpretation (VMCAI)},
 year={2016}, url={https://pmpub.inf.ethz.ch/publications/MuellerSchwerhoffSummers16.pdf}
}

@inproceedings{leino2010dafny,
 author={K. Rustan M. Leino},
 title={{Dafny}: An Automatic Program Verifier for Functional Correctness},
 booktitle={Logic for Programming, Artificial Intelligence, and Reasoning (LPAR)},
 year={2010}, url={https://www.microsoft.com/en-us/research/wp-content/uploads/2008/12/dafny_krml203.pdf}
}

@inproceedings{demoura2021lean,
 author={Leonardo {de Moura} and Sebastian Ullrich},
 title={The {Lean 4} Theorem Prover and Programming Language (System Description)},
 booktitle={Automated Deduction (CADE)}, year={2021},
 url={https://lean-lang.org/papers/lean4.pdf}
}

@inproceedings{barthe2011product,
 author={Gilles Barthe and Juan Manuel Crespo and C{\'e}sar Kunz},
 title={Relational Verification Using Product Programs},
 booktitle={Formal Methods (FM)}, year={2011},
 url={https://software.imdea.org/~jmcrespo/docs/FM2011.pdf}
}

@inproceedings{eilers2021wild,
 author={Marco Eilers and Severin Meier and Peter M{\"u}ller},
 title={Product Programs in the Wild: Retrofitting Program Verifiers to Check Information Flow Security},
 booktitle={Computer Aided Verification (CAV)}, year={2021},
 url={https://pmpub.inf.ethz.ch/publications/EilersMeierMueller21.pdf}
}

@misc{ho2022aeneas,
 author={Son Ho and Jonathan Protzenko},
 title={{Aeneas}: {Rust} Verification by Functional Translation},
 year={2022}, howpublished={arXiv:2206.07185},
 url={https://arxiv.org/abs/2206.07185}
}

@article{maciver2019hypothesis,
 author={David R. MacIver and Zac Hatfield-Dodds and others},
 title={{Hypothesis}: A New Approach to Property-Based Testing},
 journal={Journal of Open Source Software}, year={2019}, volume={4}, number={43},
 pages={1891}, doi={10.21105/joss.01891},
 url={https://joss.theoj.org/papers/10.21105/joss.01891}
}

@misc{crosshair,
 author={{CrossHair contributors}}, title={{CrossHair}: Introduction}, year={2026},
 note={Documentation accessed September 11, 2026},
 url={https://crosshair.readthedocs.io/en/latest/introduction.html}
}

@misc{basedpyright,
 author={{basedpyright contributors}}, title={{basedpyright} Documentation},
 year={2026}, note={Accessed September 11, 2026},
 url={https://docs.basedpyright.com/latest/}
}

@misc{loughridge2024dafnybench,
 author={Chloe Loughridge and Qinyi Sun and Seth Ahrenbach and Federico Cassano and Chuyue Sun and Ying Sheng and Anish Mudide and Md Rakib Hossain Misu and Nada Amin and Max Tegmark},
 title={{DafnyBench}: A Benchmark for Formal Software Verification},
 year={2024}, howpublished={arXiv:2406.08467},
 url={https://arxiv.org/abs/2406.08467}
}

@misc{banerjee2026dafnypro,
 author={Debangshu Banerjee and Olivier Bouissou and Stefan Zetzsche},
 title={{DafnyPro}: {LLM}-Assisted Automated Verification for {Dafny} Programs},
 year={2026}, howpublished={arXiv:2601.05385},
 url={https://arxiv.org/abs/2601.05385}
}

@article{kirchner2015framac,
 author={Florent Kirchner and Nikolai Kosmatov and Virgile Prevosto and Julien Signoles and Boris Yakobowski},
 title={{Frama-C}: A Software Analysis Perspective},
 journal={Formal Aspects of Computing}, year={2015}, volume={27}, number={3},
 pages={573--609}, doi={10.1007/s00165-014-0326-7},
 url={https://doi.org/10.1007/s00165-014-0326-7}
}

@misc{framacwp,
 author={{Frama-C contributors}}, title={{WP}: Deductive Proofs of {ACSL} Contracts},
 year={2026}, note={Project documentation accessed September 13, 2026},
 url={https://www.frama-c.com/fc-plugins/wp.html}
}

@misc{strata2026architecture,
 author={{Strata contributors}}, title={{Strata} Architecture},
 year={2026}, note={Project documentation, revision 89c885e59604. Accessed September 13, 2026},
 url={https://github.com/strata-org/Strata/blob/89c885e59604b22a5b63a0dc271ad40630ea02b8/docs/Architecture.md}
}

@misc{blatter2018rpp,
 author={Lionel Blatter and Nikolai Kosmatov and Pascale {Le Gall} and Virgile Prevosto and Guillaume Petiot},
 title={Static and Dynamic Verification of Relational Properties on Self-Composed {C} Code},
 year={2018}, howpublished={arXiv:1801.06876},
 url={https://arxiv.org/abs/1801.06876}
}
\appendix
\section{Historical domains and pinned revisions}
\label{app:domains}

Table~\ref{tab:domains} records the callable domains for the fixed ten-project
cohort. The source bundle contains a machine-readable copy of the complete
commit identifiers, domains, and verdicts in
\texttt{evidence/historical-scopes.json}. These records describe selected
boundaries. Passing a functional study does not broaden the corresponding
historical theorem. Exact types exclude subclasses with overridden behavior.

\begin{table}[h]
\centering\small
\caption{Declared domains and abbreviated old/new commits. Full identifiers are retained in the evidence index.}
\label{tab:domains}
\begin{tabularx}{\linewidth}{@{}l>{\raggedright\arraybackslash}p{2.4cm}>{\raggedright\arraybackslash}X@{}}
\toprule
Repository & Old / new commit & Domain and observations \\
\midrule
Black & \shortstack[l]{\texttt{bdcad23abe}\\\texttt{6409f0a51b}} & Exact two-integer tuple. Boolean return. \\[3pt]
Django & \shortstack[l]{\texttt{5cba975d26}\\\texttt{2508be35ca}} & Exact Python 3 integers. Base36 string or ValueError. Unreachable Python 2 branch removed. \\[3pt]
CPython & \shortstack[l]{\texttt{60403a5409}\\\texttt{ebf955df7a}} & Exact positive integer year. Integer result. Other calendar helpers excluded. \\[3pt]
Werkzeug & \shortstack[l]{\texttt{6389612fd1}\\\texttt{c1a26b45fb}} & Unicode-scalar string and exact Boolean weak, including omitted False. String or ValueError. \\[3pt]
Packaging & \shortstack[l]{\texttt{f585376280}\\\texttt{3b77a26f5a}} & Unicode-scalar string and Boolean validate, including its keyword-only default. String or InvalidName. \\[3pt]
python-stdnum & \shortstack[l]{\texttt{d66998e6c6}\\\texttt{52d316200c}} & Unicode-scalar number and alphabet, alphabet length greater than one, including the default. Integer or modeled exception. \\[3pt]
PyPNG & \shortstack[l]{\texttt{0dfbfda89e}\\\texttt{4632d79ce4}} & Equal-length exact bytearrays, positive filter unit. Result may alias scanline, previous distinct. Observe all buffer poststates. \\[3pt]
SGLang & \shortstack[l]{\texttt{6916ffc4cb}\\\texttt{0bcd822377}} & Stable request snapshots, committed between zero and allocated, nonnegative reserve, positive page size. Ordered lengths and total increment. \\[3pt]
PyTorch & \shortstack[l]{\texttt{08187d9e0f}\\\texttt{2b3ec34829}} & Stable ShardMetadata snapshots, equal ranks, positive sizes, integer offsets. Boolean overlap. \\[3pt]
vLLM & \shortstack[l]{\texttt{74c96922ec}\\\texttt{98dff2a81d}} & Exact integer list and integer-or-None lower\_bound, including its keyword-only default. Integer or ValueError. \\[3pt]
\bottomrule
\end{tabularx}
\end{table}

\section{Evaluation provenance and reproduction}
\label{app:reproduction}

The paper assembles retained development studies rather than a single
prospectively randomized benchmark. The claim ledger maps reported quantities
to repository-relative artifact paths and hashes. The dual-backend cohort,
the ten historical boundaries and the frozen transfer cohort are distinct
inventories. The expanded 25-unit paired cohort includes three later stdnum
units and the full CPython time parser. The historical 21-unit count is retained
only in its original evidence.

The historical comparison uses pinned Python~3.12.2, Dafny~4.11.0, and
basedpyright~1.39.10. The dual-backend checks pin Lean~4.33.1. Black's decimal
model uses Unicode~15.0.0 and a 4,300-digit integer conversion limit. Other
recorded generation environments differ and are retained per artifact.
Environment setup and compiler identity must be reproduced before comparing
results across runs.

The historical transfer protocol allows 20 seconds per Dafny obligation,
180 seconds per pair, and 10 seconds for native replay. Compiler and evaluator
files are frozen before additional history is queried. Novelty is measured
against archived function fingerprints, not against model training data.
No contract repair, additional invariants, or replacement target is permitted
after selection. Earlier failures and superseded development attempts are
retained rather than counted as successes.

\noindent\begin{minipage}{\linewidth}
Run these commands from the extracted frozen supplement root, after following
its environment setup instructions. They do not target the current development checkout.
\begin{lstlisting}[style=vpycode,numbers=none,language={}]
python reproduction/historical/replay.py --out FRESH_DIRECTORY
python case_studies/lean_hardening/run_matrix.py --out FRESH_DIRECTORY --wall 900
\end{lstlisting}
\end{minipage}

These commands require the interpreter and dependency environments described by
their study manifests. The supplementary material on \href{https://openreview.net/forum?id=dEplk1BzZm}{OpenReview} includes the manuscript sources, figures, example
listings, a provenance index, and the complete 25-unit paired-proof inputs.
The expanded reproduction directory also provides historical comparison snapshots, saved maintenance candidates, runtime source and observations. Appendix~\ref{app:reproduction-map} maps these materials to replay and fresh-experiment procedures. VeriPy is open source. The supplied source snapshot and its file manifest identify the implementation used by the paired-proof artifact. The current public development branch is not a substitute for that pinned experimental version. The expanded cohort was freshly verified on Linux during the evidence refresh.
No new GPU execution was performed. The earlier Django product was rechecked with
Dafny~4.11.0 and retained artifact hash
\par\noindent
\texttt{7be988d11c72a3f3acd525b6de5efeae893eda85e9736e89429147a596b5bc14}.
\par\noindent
The exact self-contained product is included as
\texttt{evidence/django-comparison.dfy}, together with the fresh verification log.

Repository-driven extensions also establish Dafny contracts for six CPython
calendar helpers, two Packaging name functions, four python-stdnum Luhn
functions, four PyPNG inverse filters, and three Werkzeug ETag functions.
These additions expose different semantic requirements, including module constants and
dependencies, Unicode and registered regexes, checksum traversal and custom
exceptions, and in-place buffer mutation. The extension protocol reruns all ten
functional study pipelines after changes and retains failed checkpoints.
A completed full-suite checkpoint records 1,341 passing regression tests.
These are engineering checks rather than additional case studies.

\paragraph{Native integration workloads.}
The extension studies compare generated and native executions on 146,122
calendar cases, 6,040 Packaging cases, 7,740 stdnum cases, and 3,360 PyPNG cases.
Werkzeug has 613 header cases and 11,349 regex capture/position comparisons.
Integration includes 734 guarded public date calls, 200 IMEI validations,
four complete PNG decodes, and 80 Response set/get calls. Each count belongs to
its own runner and observation scheme. Overlapping executions are not pooled
into a benchmark size. The explicit typing views also matter. CPython's resolved
initializer and Werkzeug's constructor stub are checked views, not evidence
that unchanged upstream annotations pass the strict gate.

\paragraph{AI assistance.}
AI tools assisted implementation, proof development, experiment tooling,
manuscript drafting, and figure-generation code. Figures use retained measurements.
This assistance is part of the development process, whose
total human time and model usage were not completely metered. The reported
verification and execution outcomes come from retained tool artifacts.
Appendix~\ref{app:agent-study} separately reports controlled model trials and
prospectively recorded preparation failures. Their descriptive results are tied
to each fixed protocol and do not establish general autonomous-agent effectiveness.

\section{Runtime study details}
\label{app:runtime}

\begin{figure}[ht]
\centering
\includegraphics[width=\linewidth]{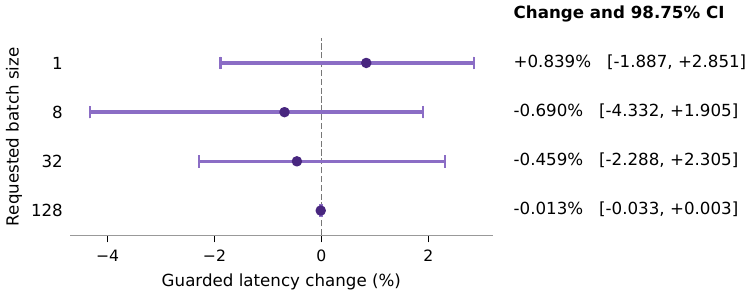}
\caption{SGLang guarded execution latency relative to native execution.
Dots show geometric-mean paired changes and bars show family-adjusted 98.75\%
BCa confidence intervals over ten matched process pairs. Positive values mean
slower guarded execution. All intervals include zero, but no equivalence margin
was specified. The narrow batch-128 interval is specific to this protocol.}
\label{fig:runtime-ci}
\end{figure}

The expanded SGLang study uses Qwen3-1.7B with an EAGLE3 draft on one A10G,
FP16, and Triton attention. CUDA graphs and prefix caching are disabled.
Each process performs two warmups and five measured generation calls per
requested batch size, with 64 output tokens per request. Ten matched process
pairs are the sampling units. Figure~\ref{fig:runtime-ci} reports the geometric
mean of paired ratios of process-median completion latencies. BCa intervals
use 20,000 resamples of complete pairs. The 98.75\% intervals adjust across
four batch sizes. No equivalence margin was specified.

The study records 400 timed calls and 16,900 requests, plus independent
mixed-length audits. Batch~128 became faster from the first to the last
measurement in all 20 processes (median $-0.324\%$), suggesting an order or
residual-warmup effect whose cause was not isolated. All calls remain included.
The narrow interval describes this protocol rather than a universal overhead
bound. Direct component timings on the same host identify deep checks and
copy-in as major boundary costs.

Black's component measurements use two warmups and nine measured batches of
30 calls with alternating measurement order. The size-32 normalizer processes
ten source lines. PyTorch's metadata measurements use 15 paired rounds after
two warmups with randomized mode order. They are CPU microbenchmarks, not
inference-latency measurements. The PyTorch integration uses Torch~2.8.0 CPU
with selected pinned helper source, a mixed-version metadata integration.

\section{Earlier Django implementation}
\label{app:old-code}

The following listing shows the checked Python~3 view of the earlier converter.
All proof annotations and executable statements are retained. The docstring
is omitted. The newer implementation appears in Figure~\ref{fig:django-example}.
Both full sidecars accompany the listings in the source bundle.
The original Django copyright and license are retained with those files.

\begin{lstlisting}[style=vpycode,language=Python]
#@ ensures raised() <==> old(i) < 0
#@ ghost_ensures not raised() ==> ghost("ExactEncoding", old(i), result)
def int_to_base36(i: int) -> str:
    digits = '0123456789abcdefghijklmnopqrstuvwxyz'
    factor = 0
    if i < 0:
        raise ValueError('Negative base36 conversion input.')
    while True:
        #@ invariant i >= 0 and 0 <= factor <= i + 1
        #@ invariant factor == 0 or 36 ** factor <= i
        #@ decreases i + 1 - factor
        #@ proof PowerPositive(factor + 1)
        factor += 1
        if i < 36 ** factor:
            factor -= 1
            break
    #@ proof StartEncoding(i, factor)
    base36: list[str] = []
    while factor >= 0:
        #@ invariant factor >= -1 and i >= 0
        #@ invariant ghost("LeadingProgress", old(i), i, factor, base36)
        #@ decreases factor + 1
        #@ proof PythonLeadingStep(old(i), i, factor, base36, digits)
        j = 36 ** factor
        base36.append(digits[i // j])
        i = i % j
        factor -= 1
    return ''.join(base36)
\end{lstlisting}

The shared \texttt{ExactEncoding} predicate fixes the result, whereas the
two versions use different progress predicates. In the newer version the
remaining quotient decreases. In the older version the first loop finds
the leading position and the second decreases that position.
The source comments are interpreted only by \toolname{} and do not invoke
Dafny code during native Python execution.

\section{Component size and proof support}
\label{app:effort}

Counts are physical nonblank lines. Python LOC excludes comments and docstrings
but includes signatures, type annotations, imports and executable assertions.
Spec counts \texttt{requires}, \texttt{ensures} and \texttt{ghost\_ensures} annotation
lines. Hint counts invariants, decreases clauses and proof hooks. Sidecar support
counts all noncomment lines, including specification definitions and reused
lemmas. It is not a count of newly authored proof lines or human effort.
The component-specific sidecars exclude imported shared libraries and generated
backend models/preambles. Shared imported Lean libraries contain 30 LOC (allocation), 19 LOC (arithmetic), 18 LOC (ranges).
Their use is recorded per unit in \texttt{evidence/component-effort.json}.
Copied support remains in each sidecar's count. No total is given because units
and their helpers overlap. Zero sidecar support does not imply zero manual work.

Times come from the same fresh Linux CI checkpoint, one public-API invocation per
unit/backend, including encoding and prover startup. Groups ran on separate CI
runners. These observations are offline costs, not a fair backend-speed ranking
or application-runtime overhead. The counting script and hashed sources are
included in the supplement.

\begin{table}[h]
\caption{All 25 paired units under the same counting policy. D / L denotes
Dafny / Lean. Every unit passed both backends.}
\centering\small
\begin{tabular}{@{}lrrrrl@{}}
\toprule
Unit & Py & Spec & Hint & Support D / L & Time D / L (s) \\
\midrule
black-mapping & 49 & 1 & 4 & 25 / 99 & 4.72 / 4.24 \\
black-normalizer & 18 & 1 & 2 & 7 / 56 & 5.14 / 3.95 \\
black-parser & 20 & 3 & 3 & 0 / 78 & 5.14 / 5.81 \\
django-base36 & 11 & 2 & 7 & 27 / 106 & 5.37 / 6.84 \\
pytorch-pair & 13 & 2 & 1 & 0 / 35 & 5.13 / 4.83 \\
pytorch-search1d & 17 & 4 & 5 & 84 / 278 & 7.51 / 5.11 \\
pytorch-searchnd & 53 & 10 & 13 & 168 / 927 & 17.27 / 66.17 \\
pytorch-validation & 94 & 18 & 27 & 357 / 1759 & 22.66 / 110.76 \\
sglang-live & 27 & 12 & 14 & 79 / 125 & 6.36 / 7.15 \\
01-pytorch-sharding & 4 & 12 & 1 & 61 / 20 & 5.05 / 0.65 \\
02-vllm-alignment & 6 & 9 & 3 & 61 / 69 & 4.69 / 2.80 \\
03-sglang-prefill-budget & 4 & 6 & 0 & 0 / 0 & 4.49 / 0.48 \\
04-cpython-rounding & 8 & 3 & 1 & 61 / 33 & 4.82 / 3.41 \\
05-cpython-calendar & 6 & 8 & 0 & 0 / 0 & 5.17 / 3.19 \\
06-sglang-first-milestone & 4 & 5 & 2 & 71 / 44 & 4.57 / 2.53 \\
sglang-aligned & 27 & 13 & 14 & 80 / 148 & 6.81 / 6.56 \\
pytorch-tensor & 33 & 3 & 10 & 16 / 236 & 6.85 / 7.26 \\
black-adjusted & 57 & 8 & 8 & 6 / 212 & 5.07 / 8.81 \\
black-consumer & 57 & 7 & 15 & 89 / 308 & 8.32 / 10.48 \\
black-composition & 98 & 8 & 19 & 119 / 421 & 8.98 / 12.48 \\
vllm-padding & 17 & 6 & 6 & 0 / 266 & 5.42 / 12.21 \\
luhn-exact & 7 & 4 & 1 & 67 / 70 & 4.16 / 1.83 \\
luhn-roundtrip & 13 & 12 & 2 & 120 / 260 & 7.49 / 2.14 \\
luhn-validation & 26 & 13 & 0 & 0 / 59 & 3.20 / 2.36 \\
cpython-time-parser & 37 & 34 & 0 & 0 / 976 & 12.73 / 163.59 \\
\bottomrule
\end{tabular}
\end{table}

\clearpage
\section{Later record-boundary development and reuse}
\label{app:records}

This experiment follows the historical checkpoint in Table~\ref{tab:historical}.
It does not rewrite that checkpoint or the failed-transfer cohort. The selected
SGLang planner and PyTorch pair-overlap bodies match their pinned old and new
upstream versions after removing annotations and docstrings. Both are unchanged
bodies. Closed snapshots expose only the accessed fields, excluding descriptors,
identity, concurrent mutation, and other upstream object behavior. Matching
schemas include field names and types. Schema migration is not supported.

The comparison generates total maps between the two modules' nominal record
values, preserving nested fields and sequence elements. Input admission is
checked on the mapped snapshot. Normal output equality is decomposed into
checked tuple/list obligations. Record field and borrowed-list mutation remain
unsupported. Functional record encodings already existed in both backends.
This extension connects them to the Dafny compatibility product, not a new Lean
relational checker or a mechanized translation-correctness theorem.

\begin{table}[h]
\centering\small
\caption{Two known record boundaries before and after the extension, with
identical supplied Python source and annotations in both runs. Both final
products prove compatible. Times are individual local wall times under
concurrent load. They do not establish a speed comparison.}
\begin{tabular}{@{}lrrrr@{}}
\toprule
Component & Py / Spec / Hint & Existing support D / L & New relation & Product (s) \\
\midrule
SGLang & 27 / 12 / 14 & 79 / 125 & 13 & 8.71 \\
PyTorch & 13 / 2 / 1 & 0 / 35 & 0 & 8.86 \\
\bottomrule
\end{tabular}
\end{table}

Both pre-extension runs reject record admission. SGLang development initially
fails three product attempts because of missing sequence-map bounds and insufficient
output-equality reasoning. A checked induction equates the independently named
sum functions. Explicit structural obligations close the product. The compiler
is frozen after SGLang development, before the PyTorch attempt. PyTorch then
passes with its existing annotations and no new relation lemma or compiler
change. Because both cases were previously known, this is cross-component reuse,
not an independently selected generalization experiment. Counts follow
Appendix~\ref{app:effort}. Existing sidecars are not newly authored effort.
Historical authoring hours and model costs are unknown.

Fresh functional checks pass in both Dafny and Lean for both components.
Two synthetic behavioral mutations yield native witnesses by reading committed
instead of allocated length in SGLang, and changing the touching-boundary test
in PyTorch. A retained PyTorch witness uses nonnegative offsets and positive
sizes. Three mutation/alias controls are rejected by each backend (six checks).
A deliberately corrupted field map also produces a detected proof/replay
contradiction in regression testing. The targeted suite passes 138 tests.
These controls do not establish general compiler soundness.

The supplement includes separate pre-extension and final compiler snapshots,
source/contract bindings, relation support, observations, and replay commands.
Both products replay successfully from the final snapshot. This later local
experiment is separate from the 50-invocation Linux functional checkpoint.

\section{Agent experiments and preparation coverage}
\label{app:agent-study}

\subsection{Task and information boundaries}

Initial verification asks an agent to supply proof annotations and checked
lemmas for fixed Python source and requirements. Maintenance additionally asks
for old-domain inclusion and equality of declared observations after an update.
Both conditions receive the same executable old and new sources, public
specifications, domain, syntax guide and checker. Only the old proof hints and
task-specific checked helpers differ. Source, types, contracts and input
restrictions cannot be changed during proof attempts. Reuse and rebuild trials
have fresh sessions. Product diagnostics that might reveal evaluator-private
old proof material are withheld from both conditions.

The two development pilots each contain ten trials on the same Black and Django
components. Their old proofs were prepared during earlier development with the
after version already known, and that preparation cost is unknown. The first
pilot required a second response for each Django rebuild trial because loop
invariants were placed outside the loop body. After the common guide explicitly
showed legal placement inside a loop, all ten trials in the second pilot
completed on their first response. This removes the observed response-count
difference and does not support a general reuse advantage.

A separate measured old-only stage supplies fixed requirements but removes the
reference proof and after version from the author packet. Black succeeds in one
response. Django uses all three responses without a checked proof. The first
candidate encounters a division precondition and lemma timeouts, the second uses
a disallowed declaration, and the final candidate times out on two lemmas.
Black's proof is sealed to its source, requirements, candidate, artifacts and
implementation before a saved update candidate is replayed successfully.
That replay uses no new model call. Django receives no maintenance packet.
This experiment validates the stage on known requirements and does not make the
earlier requirement design prospectively blinded.

\subsection{Prospective coverage protocol}

The later study freezes the complete executable protocol locally before
successful acquisition. The ten repository and file identities and pinned
ancestry roots remain those chosen during development. Each repository has
three slots. Selection inspects at most ninety path-history commits in the
archived API order, compares each with its first parent, enumerates common
top-level functions alphabetically, and retains the first eligible change per
function name until three slots are filled. The executable body must change,
and its after-body fingerprint must be absent from the refreshed development
inventory. Documentation, signature types and inert literal exception-message
changes alone do not qualify. There is no replacement after preparation results.
Novelty is relative to archived development, not model training data.

The old-only preparation process starts with the static declaration closure.
It permits insertion of missing builtin signature annotations and documented
contracts while retaining every original executable byte and existing type.
One model response proposes a requirement and three to eight concrete examples.
A separate fresh-session response reviews any valid proposal against the old
documentation. Undocumented restrictions, vacuity and replacement object
interfaces are disallowed. This proposed review is model-assisted and would
not constitute independent human certification. In the observed run, every
proposal author abstained, so no proposal reached that review step.

Admitted old tasks first receive a baseline check, then at most three proof
responses if necessary. Only a checked old proof can release an update. The
update policy preserves the frozen old functional property where signature
layouts permit type and contract transfer. Eligible tasks receive one initial
property trial and two maintenance repetitions per arm, each capped at three
responses. Thus the thirty-slot maximum is 600 task calls and one isolation
call. The model is \texttt{gpt-6-astra} with medium reasoning, a 300-second model
wall, a 180-second checker wall and 20 seconds per Dafny obligation. The sample
is resource-bounded and does not have a power-based efficacy justification.

\begin{table}[ht]
\centering\small
\caption{Prospective preparation outcomes. No selected change reaches a reviewed
contract or proof trial. Unfilled slots remain part of the thirty-slot frame.}
\label{tab:preparation-coverage}
\begin{tabular}{lrrrr}
\toprule
Repository & Selected & Unfilled & Closure blocked & Author abstained \\
\midrule
Black & 0 & 3 & 0 & 0 \\
Django & 3 & 0 & 3 & 0 \\
CPython & 1 & 2 & 1 & 0 \\
python-stdnum & 0 & 3 & 0 & 0 \\
Werkzeug & 3 & 0 & 3 & 0 \\
Packaging & 3 & 0 & 2 & 1 \\
PyPNG & 3 & 0 & 0 & 3 \\
SGLang & 3 & 0 & 0 & 3 \\
PyTorch & 0 & 3 & 0 & 0 \\
vLLM & 3 & 0 & 1 & 2 \\
\midrule
Total & 19 & 11 & 10 & 9 \\
\bottomrule
\end{tabular}

\end{table}

\subsection{Outcomes, limitations and integrity}

Static extraction reports multiple bindings, retained dependency mutation,
conservative checks on module calls and an unresolved module identity. These
are preparation barriers rather than demonstrated backend rejections.
The nine requirement authors explicitly return \texttt{not-prepared}, citing
insufficient documented behavior or missing custom-object and dependency
semantics. The builtin-type and JSON-example preparation interface itself limits
object-oriented tasks. Zero coverage under this policy does not imply that no
manual VeriPy specification or adapter could verify those components.

No selected task reaches strict type checking, a backend, native observation,
proof sealing or maintenance. Proof-generation accuracy and reuse effects are
therefore not estimable. The successful acquisition run uses nine task calls
and one isolation call, reporting 141,274 input tokens, including 77,568 cached
input tokens, and 1,505 output tokens. Actual billed cost is unavailable.
An earlier TLS-failed acquisition retains its separate isolation call and all
ten retrieval failures. No repository contents were received in that run.

The successful protocol binds a verified system CA configuration before
acquisition. The audit replays selection, reconstructs all nine task prompts,
checks ten distinct sessions and verifies 322 response hashes. The model cannot
read the repository, use recorded tools or retrieve the update. A live canary
and OS denial check pass. Nineteen pre-freeze harness and selection tests pass,
including a synthetic handoff with the real checker and mocked model responses.
Six targeted assurance controls also pass. The synthetic fixture is not a
repository task or a model performance observation.

The artifact retains an initial environment-recording failure and a post-run
auditor correction. The original offline auditor represented an unavailable
source as \texttt{FileNotFoundError} while acquisition recorded \texttt{OSError}.
A separately hashed auditor corrects that replay category and byte decoding.
It does not change the frozen selector, source inputs or experiment outcomes.
The compiler is unchanged from the previous all-ten passing checkpoint.

\subsection{A subsequent evaluation under a shared frozen policy}
\label{app:shared-policy}

A subsequent run separates matched maintenance trials on known components from
prospective preparation coverage. Both arms use the same strict
\texttt{basedpyright} policy, audited regex interface, compiler, public
specifications, three-response budget and 240-second checker limit. The model
remains \texttt{gpt-6-astra} with medium reasoning. Old proof comments and
checked task-specific helpers are available only to the reuse arm. Both arms
receive the same restricted functional feedback. Product diagnostics are
withheld because they can contain private old proof material.

A human project collaborator reviewed the six known requirements, clarified
the permissive list-header scope, and approved a corrected checksum example.
The initial incorrect example and rejection remain in the record. This is a
separate human assessment of agent-prepared requirements, not external
certification of specification adequacy.

The planned known-case schedule contains twenty-four trials. Packaging has a
review-ID binding failure before dispatch, leaving four slots unexecuted. The
operator continues the other individually approved tasks through the unchanged
trial and checking functions. This taskwise continuation is a recorded
administrative deviation from the entry point that gates the entire schedule.
It changes neither the per-trial budget nor the candidate acceptance rules.

Twenty trials execute. The known Django algorithmic rewrite succeeds in both
repetitions of both arms. Each arm uses three responses in total across its two
repetitions. Thus this case provides no observed completion or response-count
advantage from old proof access. The remaining executed tasks use unchanged
source and must not be pooled with Django as real maintenance changes.
List-header and checksum reuse trials establish functional proofs but not
compatibility. Their rebuild trials establish neither joint obligation.
All twenty-four quoting and unquoting responses alter blank-line counts while
preserving executable ASTs and nonblank lines. The frozen exact-text rule rejects
these submissions before theorem checking. These outcomes measure a submission
interface failure rather than theorem difficulty or behavioral incompatibility.
The ETag controls also have no task-specific old proof support contrast.

The same run registers twenty prospective slots across the ten repository paths
before acquisition. Fourteen changed bodies across seven repositories are absent
from the frozen exposure inventory, and six slots remain unfilled. Eleven
requirements proposals are rejected and one dependency closure is blocked.
Two model-reviewed proposals reach the human reviewer. The CPython time-parser
proposal admits two mixed-separator inputs rejected by the old implementation,
and the human reviewer rejects it. The Django ETag proposal is retained only as
a desired partial property. Its all-Python-strings domain exceeds the current
Unicode-scalar boundary, and its baseline fails strict typing at an imported
helper. No selected task obtains an adequate old proof or enters maintenance.
The conditional unseen maintenance effect remains unestimable.

The run records sixty-six model calls, comprising fifty maintenance responses,
fifteen requirements proposal or review responses, and one isolation response.
All transcripts reparse with distinct sessions and zero recorded tool activity.
Forty-six checker receipts pass their artifact hash audit. Twenty integrity tests
pass, and a fresh archive extraction reproduces all four Django proof successes
without model calls. The archive preserves every selected failure, review
decision and administrative deviation. These checks establish the reported
reproduction scope, not correctness of the trusted frontend or independence
from model training data.

\subsection{Preparation repair and a second frozen maintenance policy}
\label{app:maintenance-v2}

The revised maintenance interface replaces regenerated source with line-addressed
proof edits. The evaluator reconstructs each proof proposal against immutable
statement locations and checks both executable AST and nonproof bytes. An
explicit identifier map binds each task to its exact reviewed requirement.
Packaging's formal predicate is strengthened to require the exception class
\texttt{InvalidName}, matching its approved prose. The human collaborator
approves this correction before the new study. The implementation also extends
the compatibility boundary to supported optional values. The all-ten
checkpoint passes all fourteen jobs and 98 regression checks.

Two frozen development pilots exercise the revised interface. The first
establishes ten joint proofs in twelve trials over six known components.
Five components are identity controls and Django is the known rewrite.
After clarifying permitted ghost annotations, a second pilot establishes
two joint proofs in four identity-control trials on list-header and checksum.
Both rebuild trials exhaust their budgets. These development-selected results
guide preparation and are not unseen maintenance measurements. Full-module
preparation controls subsequently pass strict typing and existing reference
proofs for all six components.

\paragraph{Natural-history continuation.}
A new registration reserves twelve slots over six function families in four
repositories. It freezes the compiler, preparation, selection rule and
7,141-body exposure inventory before acquisition. Newest-first pinned path
history supplies one Werkzeug unquoting change and two Django base36 changes.
Nine slots remain unfilled. Selection uses changed executable bodies and
archival novelty, with no outcome-based replacement.

The collaborator delegates the three old-requirement decisions to the
authoring model. We therefore retain the original direct-human-review
registration and label the subsequent runs exploratory protocol amendments.
The model approves Werkzeug's Unicode-scalar, permissive parsing requirement.
It rejects both Django proposals because the old code violates the requested
Python~3 behavior. One reads the absent \texttt{sys.maxint} attribute on
nonnegative inputs. The other returns \texttt{z} for $-1$ instead of raising
\texttt{ValueError}. These decisions are not independent human review.

Werkzeug preparation first duplicates contracts on typing overload stubs.
Six old-author responses fail. Removing those clauses exposes unsupported
duplicate definitions, and six further responses fail. A third amendment
projects recognized typing-only stubs while retaining the final callable AST
and annotations. Original overload declarations pass strict typing separately.
Typing-registry and module-introspection observations are excluded. The old
baseline then proves without task-specific hints and is sealed before update
release. These are three continuations of one cohort, not three new samples.

All four maintenance trials exhaust three responses. All twelve proposals
pass the source-edit interface, but the new implementation uses unsupported
start-offset string methods. No joint proof or formal incompatibility result
is obtained. Separately, native execution of the pinned bodies confirms that
a single-space input changes from an empty returned tag to a single-space tag.
This post-release witness is distinct from the unsupported proof verdict.
The old baseline has no task-specific support, so the two packets also lack a
proof-reuse treatment contrast. The natural-history result cannot estimate a
reuse effect.

\paragraph{Controlled authored changes.}
A separate registration seals the six directly human-reviewed old packets
and their checked reference proofs before generating any update. Twelve
predeclared transformation requests cover branch restructuring, string
construction, indexed traversal, checksum parity and base36 conversion.
Each request receives one isolated author response. The author sees the old
source, public contract and request, but no private old proof support.
All twelve proposals are admitted. Admission requires an executable change
absent from both raw and normalized archived bodies. Normalization removes
local renaming, local type annotations and literal exception-message changes.
No typing, proof, native agreement or compatibility result determines admission.

These updates are generated for the experiment and are not upstream commits.
The same model family serves separate author and maintainer sessions.
The design reserves two repetitions per arm and three responses per trial,
with arm order counterbalanced by transformation and repetition. Both arms
use \texttt{gpt-6-astra} at medium reasoning, identical source and contracts,
public relation libraries, strict typing and checker limits. Only checked old
proof comments and task-specific helpers differ. Rebuild submissions cannot
refer to private old helpers, including guessed names. Eight transformations
have nonempty old support. Four quoting and unquoting transformations are
equally informed controls and remain in the full denominator.
The checker receives the same verified old model in both conditions. The
treatment changes the maintainer's access to and permitted references to old
proof material.

The old function bodies contain 5 to 17 Python lines and the generated bodies
7 to 14, counting the definition and executable nonblank lines but excluding
docstrings and comments. Public fixed specifications contain 3 to 19 nonblank
sidecar lines. The four components with private old helpers supply 25 to 148
nonblank lines, and the shared relation material contains 11 to 37. These
sidecar counts include comments and exclude the common compiler's semantic
libraries. Manual specification, library and old-proof authoring time was not
prospectively logged and is unknown. Maintenance response counts do not measure
that setup cost.

Joint functional and compatibility proof is the primary endpoint. Completion,
unsupported outcomes and exhausted budgets are reported separately from
response costs. Repetitions are nested within transformations, functions and
repositories. Finite post-generation native checks agree for all twelve
updates, but are neither proofs nor model feedback. Public specification and
relation development, manual old-proof preparation and collaborator review
remain part of the disclosed setup effort.

All 48 primary trials complete. Reuse establishes 14 joint proofs with 34
responses, and rebuild establishes ten with 46. All eighty maintenance
submissions pass the immutable-source interface. The paired outcomes comprise
ten repetitions where both succeed, four reuse-only successes and ten where
neither succeeds. Every reuse-only success belongs to Packaging. Among the ten
pairs with joint completion in both arms, response counts total twelve for
reuse and sixteen for rebuild. Seven pairs have equal response counts and
three use fewer responses with reuse. Repetitions remain clustered observations.
At the transformation level, reuse verifies seven of twelve and rebuild five,
with completion consistent across both repetitions for every arm and target.

Twelve trials stop at strict Python typing because the two header updates and
one base36 update omit local list element annotations. Four quoting trials
encounter unsupported string augmented assignment. Four unquoting trials emit
ill-typed optional-string conjunctions and indexing in Dafny. Their raw
\texttt{failed}/\texttt{inconclusive} statuses do not denote unresolved
mathematical obligations. Of the four unsuccessful Packaging rebuild trials,
three end with unproved obligations and one with a proposed-model resolution
error. The raw diagnostics remain available alongside the descriptive categories.

The primary study records 93 model invocations, comprising twelve change-author
calls, eighty maintenance responses and one isolation probe. All have distinct
sessions and zero recorded tool activity. The audit reconstructs the complete
prompts, histories, feedback and accepted candidates, checks all eighty checker
receipts, and validates the old proof seals and exact review bindings. Recorded
usage is 1,268,676 input tokens, including 752,384 cached tokens, and 101,079
output tokens. The provider separately reports 23,401 reasoning output tokens.
These fields may overlap and actual billing is unavailable.

\paragraph{Public library information diagnostic.}
The Packaging old proof contains a lemma relating the public requirement's
separator state machine to the encoder's recursive regex substitution model.
It also exposes the names and structure of the backend operations. Those
definitions are absent from the primary rebuild packet. Thus old proof access
bundles a worked proof with information about the compiler's API.

After completing and auditing all primary trials, we freeze a diagnostic on
both previously evaluated Packaging transformations. Both arms receive the
exact public regex preamble and the documented substitution lowering. The
source, specification, checker, model, feedback and three-response budget remain
fixed. No private old lemma is added to rebuild. Two repetitions per arm yield
eight trials, counterbalanced in the same manner as the primary study.

Both arms establish all four joint proofs in their first response. The primary
Packaging result of four reuse successes and zero rebuild successes therefore
does not persist under the added public information. All nine diagnostic
invocations, including isolation, have distinct sessions and zero recorded tool
calls. The prompt, candidate, source and receipt audit passes. This follow-up
was selected because of the observed primary gap. It is a post-hoc sensitivity
analysis on one function, not a fresh cohort or replacement score, and it does
not isolate a general causal effect of lemma reuse.

\paragraph{Fresh artifact replay.}
A fresh extraction validates all 12,745 archived file hashes and reruns the
three completed maintenance audits. The primary and diagnostic result reports
regenerate byte for byte. In a new Python environment on the same host, 36
distinct checker bundles reproduce all 47 retained check outcomes, covering
all 32 primary and diagnostic joint-proof successes, six old references, the
exploratory old baseline and eight retained failures. Identical bundles are
checked once and mapped to every corresponding original receipt. No model
calls are made. Original and replay logs are preserved separately. This checks
saved-candidate reproducibility under the recorded toolchain, not independent
external replication or correctness of the trusted frontend.

\subsection{Subsequent calendar and string development}
\label{app:postfreeze-development}

The following work occurs after the frozen maintenance experiments. It retains
their compiler snapshots, review decisions, failed attempts and reported scores.
The collaborator authorizes review by the authoring model. This permits the
exploratory continuation but does not constitute independent human review.
No additional agent maintenance calls are made in these controls or the calendar
search, and none of their outcomes changes Table~\ref{tab:controlled-maintenance}.

\paragraph{Calendar requirement and readiness.}
The model approves the exact prepared CPython \mbox{\code{\_ord2ymd}} requirement.
For exact positive integer ordinals, it requires a valid positive-year Gregorian
date whose mathematical ordinal equals the input. The requirement has no
year-9999 ceiling. The selected helper closure and resolved initializer view
remain explicit, with invalid inputs, resource failures, introspection and
concurrent mutation outside scope. Finite native checks agree with the C-backed
date oracle for all 3,652,059 public date ordinals. Eighty additional checks use
400-year periodicity at large cycle boundaries, including 4096-bit cycle counts.
These corroborate the review but do not prove the unbounded contract.

The saved old proof passes again. A checked public uniqueness relation supports
both an identity product and a scoped CPython 3.13.0-to-3.14.0 product. Both
products pass. The latter prepares the update from its pinned full module,
retaining the selected callable and required helpers. Its entry body is
unchanged. These controls establish readiness within the declared boundary,
not changed-code maintenance or whole-release compatibility.

\paragraph{Frozen calendar history search.}
A separate registration freezes a 7,436-body exposure inventory, selection and
preparation rules, compiler, delegated review authority, trial budgets, and
identical public Dafny and calendar relation material for both arms before
acquisition. It reserves one slot on each of the current and former datetime
module paths. At one pinned CPython HEAD, newest-first history compares each
revision with its first parent. Selection requires an archive-novel changed
\mbox{\code{\_ord2ymd}} body, with deduplication across paths and no typing or proof
success filter. The bound is two pages of fifty entries per path.

The search examines 36 entries on \code{Lib/\_pydatetime.py} and 98 on
\code{Lib/datetime.py}. It selects no eligible changed bodies and retains both
unfilled slots. All 262 requests, three retrieval failures, 255 stored source
files and selection decisions are preserved. Two paths represent one function
family in one repository, not independent projects. The result supplies no
maintenance comparison and does not exhaust other branches, dependency changes
or function families. Previously exposed calendar controls remain separate.

\paragraph{String offsets and a known behavioral difference.}
The later Dafny extension admits positional integer start and end bounds for
\code{startswith} and \code{endswith}, including literal tuples of strings.
Negative bounds count from the end. Unlike slicing, a positive start is not
clamped to the string length, so an empty affix fails beyond the end or in an
inverted interval. Bounds are evaluated even for an empty tuple. The existing
Unicode-scalar boundary remains. Noninteger and keyword bounds and Lean string
support are outside this extension.

The focused suite passes 339 tests, including nine affix tests with 28,285
native/generated value comparisons, symbolic interval properties and rejected
safety violations in bounds. The all-ten checkpoint passes fifteen jobs with
no source drift, including 98 existing regression tests and the nine affix tests.
Repeated checks are engineering evidence, not additional study units.

On the already exposed Werkzeug update, the old proof passes again and the new
body clears the earlier start-offset encoding barrier. The unchanged old
specification refers to the public Unicode stripping model, whose automatic
inclusion was triggered by the old body's strip operation. The update removes
that operation, leaving an unresolved model reference. A separate diagnostic
supplies the exact public Unicode definitions as checked sidecar support,
without changing either executable body or the fixed requirement. This removes
the resolution error and leaves two source postcondition failures.

Native execution of \code{unquote\_etag(" ")} returns \mbox{\code{("", False)}} in
the old version and \mbox{\code{(" ", False)}} in the update. This witnesses a
behavioral difference and is a native counterexample to compatibility.
The product-proof attempt remains inconclusive, and no backend proof of
incompatibility is claimed. The diagnostic demonstrates why
coverage repair must be separated from successful maintenance of a fixed
contract. Both attempts and their dependency assumptions remain in the artifact.

\subsection{Earlier automatic-transfer policy}

Before querying additional history, the transfer experiment freezes the
compiler, selection rule, adaptation policy, and an inventory of 1,632
function-body fingerprints from 598 source contents. For each of the ten fixed
repository/file pairs, it examines at most 30 path-history entries, compares
each commit with its first parent, and selects the first eligible changed
top-level function whose after-body is absent from that inventory.
The adapter transfers existing types only where parameter layouts match.
It adds an exact postcondition for a call-free single return, or
\texttt{ensures True} otherwise, with no new invariants, lemmas, dependency
models, input restrictions, or case-specific repairs.

Nine changes are selected. All nine are unsupported before proof. Eight first
blockers concern dependency extraction and one concerns defaults. Black supplies
no eligible novel body in the acquired history window, so its selection failure
remains in the ten-project denominator. No successful generalization follows
from this experiment. It measures a strict automatic transfer policy, not the
feasibility of manually annotating those functions. Subsequent support added
after inspecting these cases is development work and cannot turn their reruns
into fresh held-out results.

\clearpage

\subsection{Additional descriptive maintenance metrics}

\label{app:maintenance-metrics}

These metrics describe the same 48 maintenance trials, not additional experiments. Each arm has 24 trials on 12 updates, with two repetitions per update. Costs cover maintenance responses only, excluding change generation, isolation, old-proof preparation, and the later library diagnostic.

\begin{table}[ht]
\centering\small
\caption{Twelve agent-authored updates, each repeated twice per arm. Joint proof means functional correctness and compatibility. First response counts joint proofs completed after one model response. Asterisks mark controls without private old proof support. Responses include failures.}
\label{tab:controlled-maintenance}
\begin{tabular}{@{}lrrrrrr@{}}
\toprule
& \multicolumn{2}{c}{Joint proofs} & \multicolumn{2}{c}{First response} & \multicolumn{2}{c}{Responses} \\
Component & Reuse & Rebuild & Reuse & Rebuild & Reuse & Rebuild \\
\midrule
Packaging name & 4/4 & 0/4 & 4/4 & 0/4 & 4 & 12 \\
Werkzeug quote$^*$ & 2/4 & 2/4 & 2/4 & 2/4 & 8 & 8 \\
Werkzeug unquote$^*$ & 2/4 & 2/4 & 2/4 & 2/4 & 8 & 8 \\
Werkzeug list-header & 0/4 & 0/4 & 0/4 & 0/4 & 4 & 4 \\
stdnum checksum & 4/4 & 4/4 & 2/4 & 0/4 & 6 & 9 \\
Django base36 & 2/4 & 2/4 & 2/4 & 1/4 & 4 & 5 \\
\midrule
Total & 14/24 & 10/24 & 12/24 & 5/24 & 34 & 46 \\
\bottomrule
\end{tabular}
\end{table}

\begin{table}[h]

\centering\small

\caption{Completion, effort and final failure categories for the controlled study.}

\begin{tabular}{@{}lrr@{}}
\toprule
Metric & Reuse & Rebuild \\
\midrule
Joint proof rate & 58.3\% & 41.7\% \\
Updates proved in both repetitions (of 12) & 7 & 5 \\
Joint proofs within one response (of 24) & 12 & 5 \\
Joint proofs within two responses (of 24) & 14 & 9 \\
Joint proofs within three responses (of 24) & 14 & 10 \\
Total model responses & 34 & 46 \\
Mean responses per trial & 1.42 & 1.92 \\
Responses on 10 jointly successful pairs & 12 & 16 \\
Summed model-call seconds & 1,346.9 & 2,011.8 \\
Median model-call seconds per trial & 38.6 & 52.7 \\
Summed checker seconds & 811.6 & 1,453.8 \\
Reported input tokens & 485,813 & 629,500 \\
Reported cached-input tokens & 278,912 & 363,008 \\
Reported output tokens & 39,824 & 59,792 \\
Reported reasoning tokens & 6,316 & 17,085 \\
Actual dollar billing & Unknown & Unknown \\
\midrule
Failure from strict typing & 6 & 6 \\
Failure from encoding/specification & 2 & 2 \\
Failure from backend typing/resolution & 2 & 3 \\
Failure from unproved obligation & 0 & 3 \\
\bottomrule
\end{tabular}

\end{table}

Of 24 paired repetitions, 10 succeed in both arms, 4 only with reuse, none only with rebuild, and 10 in neither. These correspond to five, two, zero and five distinct updates. Repetitions agree on completion and are not independent samples.

Time totals sum individual call durations, not experiment makespan or pure solver time. Failed trials may stop at strict typing or exhaust the response budget. Their lower costs do not indicate better verification. Common-success costs are conditional on both arms completing. Reported token fields can overlap, particularly reasoning and output fields. They are not added into a billed total.

The separately frozen public-library diagnostic supplies regex definitions equally. Both arms prove 4/4 Packaging trials in one response each. The primary completion gap and these response metrics therefore do not isolate a general proof-reuse advantage. No significance test or broader generalization follows from this small, clustered cohort.

\clearpage
\section{Reproduction map}
\label{app:reproduction-map}

The \href{https://openreview.net/attachment?id=dEplk1BzZm\&name=supplementary_material}{supplementary material on OpenReview} provides an explicit path from each main result to
its source, protocol and observations. The guide in
\code{reproduction/README.md} gives environment setup, exact commands and expected
outcomes. A manifest identifies every packaged file. A separate export index
maps original hashes to the anonymized exports retained from the review artifact.
Metadata redacted in those exports has new hashes and
does not constitute a fresh validation of an original freeze receipt.

\paragraph{Functional verification and historical compatibility.}
The paired-proof artifact retains the 25 annotated units, sidecars and Linux
compiler reference. Its matrix uses Python~3.12.2, Dafny~4.11.0 and Lean~4.33.1.
The historical reproduction directory separately contains the ten comparison
requests, old and new closures, relations and archived compiler for
Table~\ref{tab:historical}. Its replay applies the original obligation and pair
budgets and retains inconclusive and unsupported outcomes. Packaging's three
native witnesses are checked from the pinned closures.

\paragraph{Maintenance and prospective selection.}
The maintenance directory supplies both arms' frozen protocols, task material,
prompts, responses and decisions. An isolated compiler and checker replay 36
distinct saved bundles covering 47 original checker outcomes, including all
primary and diagnostic joint-proof successes. Direct aggregation of retained
trials reconstructs the completion and response counts. These operations do
not resample model outputs. Fresh trajectories require access to the specified
model and the recorded session, information and stopping rules. Model sampling
is not assumed deterministic. Prospective selection records retain pinned
revisions, exposure inventories and unfilled or rejected decisions. Querying
current upstream history instead would define a different cohort.

\paragraph{Runtime integration.}
The runtime directory contains the executed SGLang source, pinned model
revisions, serving configuration, process observations and analysis code.
NumPy~2.3.2 and SciPy~1.16.3 reproduce the retained estimates and bootstrap
intervals. The analyzer checks completeness, pairing, configuration and output
agreement before computing statistics. Fresh serving measurements require an
A10G host, compatible Linux and CUDA setup, the pinned packages and model
weights. Host preparation and measurement commands accompany the source.
Statistical replay does not reproduce GPU execution or imply identical timing
on another machine.

\paragraph{Scope.}
This information supports checking the reported main results under their
stated domains and protocols. It does not claim an automatic reconstruction of
every exploratory development attempt, external service state or unmeasured
authoring cost. Earlier engineering records remain identified through the
original evidence index. The source snapshot distributed with the supplement, rather than the current
public development branch, is the version reference for the reported experiments.
Current code and documentation are available from
\url{https://github.com/astrio-labs/veripy} and
\url{https://astrio-labs.github.io/veripy/}.

\end{document}